\documentclass[fleqn,usenatbib,letters]{mnras}

\usepackage[T1]{fontenc}
\usepackage{ae,aecompl}

\usepackage{graphicx}
\usepackage{amsmath}
\usepackage{amssymb}

\title{What can we really infer from GW 150914?}

\author[J.~F.~Rodriguez et al.]{J.~F.~Rodriguez,$^{1,2}$ J.~A.~Rueda,$^{1,2,3}$ R.~Ruffini$^{1,2,3}$
\\
$^1$Dipartimento di Fisica and ICRA, Sapienza Universit\`a di Roma, P.le Aldo Moro 5, I--00185 Rome, Italy\\
$^2$ICRANet, Piazza della Repubblica 10, I--65122 Pescara, Italy\\
$^3$ICRANet-Rio, CBPF, Rua Dr.~Xavier Sigaud 150, Rio de Janeiro, RJ, 22290--180, Brazil
}

\date{Accepted XXX. Received YYY; in original form ZZZ}

\pubyear{2016}

\begin{document}
\label{firstpage}
\pagerange{\pageref{firstpage}--\pageref{lastpage}}
\maketitle

\begin{abstract}
We analyze the event GW 150914 announced by the Advanced Laser Interferometer Gravitational-Wave Observatory (LIGO) as the gravitational-wave emission of a black-hole binary merger. We show that the parameters of the coalescing system and of the newly formed Kerr black-hole can be extracted from basic results of the gravitational-wave emission during the inspiraling and merger phases without sophisticated numerical simulations. Our strikingly accurate estimates are based on textbook formulas describing two different regimes: 1) the binary inspiraling analysis treated in Landau and Lifshitz textbook, and 2) the plunge of a particle into a black-hole, treated in the Rees-Ruffini-Wheeler textbook. It is stressed that in order to infer any astrophysical information on the masses of the system both regimes have to be independently and observationally constrained by LIGO, which does not appear to be the case.
\end{abstract}


\section{Introduction}\label{sec:1}

It was recently announced by the Advanced Laser Interferometer Gravitational-Wave Observatory (LIGO) the observation, for the first time, of a gravitational wave signal \citep{2016PhRvL.116f1102A}. After the data analysis using numerical relativity templates of black-hole binary mergers, they concluded that the signal, which lasts 0.2~s with increasing gravitational-wave frequency from 35 to $\sim 150$~Hz, was emitted during the inspiral and merger of a binary black-hole, followed by the subsequent ringdown phase of the newly formed black-hole. The binary black-hole parameters obtained from this analysis are: $m_{\rm BH,1} = 36^{+5}_{-4}~M_\odot$, $m_{\rm BH,2} = 29^{+4}_{-4}~M_\odot$, and a luminosity distance to the source $d_L = 410^{+160}_{-180}$~Mpc (cosmological redshift $z=0.09^{+0.03}_{-0.04}$), adopting a flat $\Lambda$-cold-dark-matter cosmology with Hubble parameter $H_0=67.9$~km~s$^{-1}$~Mpc$^{-1}$, and matter and dark-energy density parameter $\Omega_m = 0.306$ and $\Omega_\Lambda = 0.694$, respectively.
%

We show in this Letter that the general features of the system that generates GW 150914 can be inferred from a simple analysis of the gravitational-wave signal on the light of the foundations of the gravitational-wave theory and from two classic results: 1) the analysis of the inspiraling phase of a binary system of point-like particles, and 2) the merger analysis based on the approximation of a test-particle falling into a black-hole. It is of course essential for the validity of this approximation that both regimes be independently confirmed by observations.

\section{Binary evolution}\label{sec:2}

\subsection{Inspiral phase}\label{sec:2a}

A binary system, by emitting gravitational-waves, evolves through two different phases: the regime I is the ``adiabatic'' inspiral phase in which the binary follows quasi-circular orbits. The gravitational-wave energy spectrum in this phase can be estimated from the traditional formula of the quadrupole emission within the classic point-like approximation \citep{1963PhRv..131..435P,1964PhRv..136.1224P,1974bhgw.book.....R}
\begin{equation}\label{eq:dEdfinspiral}
\frac{dE}{df} = \frac{1}{3}(\pi G)^{2/3} \nu M^{5/3} f^{-1/3},
\end{equation}
where $\nu \equiv \mu/M$ is the so-called symmetric mass-ratio parameter, with $\mu=m_1 m_2/M$ the binary reduced mass, $M=m_1+m_2$ the total binary mass, and $f$ is the gravitational-wave frequency. We recall that $f = 2 f_s = 2 f_{\rm orb}$, where $f_s$ the source frequency so the orbital one, i.e.~$f_s=f_{\rm orb} = \omega_{\rm orb}/(2\pi) = \sqrt{G M/r^3}/(2\pi)$ and $r$ the binary separation distance. We recall that the quantity $M_{\rm chirp}\equiv \nu^{3/5} M$ is referred in the literature to as the binary chirp mass.

When the conservative quasi-circular dynamics following the classic point-like approximation breaks-down, the regime II, namely the final plunge, merger and ringdown of the newly formed object sets in. We denote the gravitational-wave frequency at which the quasi-circular evolution ends as plunge starting frequency, $f_{\rm plunge}$. We adopt here as an estimate of the plunge starting frequency, the one of the last stable orbit (LSO) of a test-particle around a Schwarzschild black-hole:
\begin{equation}\label{eq:fLSO}
f_{\rm plunge}\approx f_{\rm LSO}=\frac{c^3}{G} \frac{1}{6^{3/2} \pi M} \approx 4.4 \frac{M_\odot}{M}~{\rm kHz}.
\end{equation}
Thus, the total energy radiated during the inspiral regime can be estimated from the binding energy of the LSO which, by extrapolation to the case of a binary of comparable masses, reads $\Delta E_{\rm inspiral} = (1-\sqrt{8/9}) \mu c^2$, where we recall that $\mu$ is the binary reduced mass. Thus, for a symmetric binary it gives $\Delta E_{\rm inspiral} \approx 0.014 M c^2$.

\subsection{Plunge, merger and ringdown}\label{sec:2b}

After the regime I of quasi-circular inspiral evolution, the regime II composed by the plunge, merger and ringdown, as first analyzed in \citet{1971PhRvL..27.1466D,1972PhRvD...5.2932D} for a test-particle falling radially into a Schwarzschild black-hole, starts. It was shown first by \citet{1971PhRvL..27.1466D,1972PhRvD...5.2932D} that the gravitational-wave spectrum in this regime II is dominated by the $l=2$ multipole (quadrupole) emission and that the largest gravitational-wave emission occurs from $r\approx 3 G M/c^2$, at the maximum of the effective potential
%
\begin{align}
&V_l(r) = \left( 1- \frac{2 m_{\rm BH}}{r} \right) \times \\
&\left[\frac{2 \lambda^2 (\lambda+1) r^3 + 6 \lambda^2 m_{\rm BH} r^2 + 18 \lambda m_{\rm BH}^2 r + 18 m_{\rm BH}^3}{r^3 (\lambda r + 3 m_{\rm BH})^2} \right]\nonumber
\end{align}
%
where $\lambda = (l-1)(l+2)/2$, and the black-hole horizon. It was there shown that in the limit of large $l$, the contribution of each multipole to the spectrum peaks at the gravitational-wave frequency
\begin{equation}\label{eq:fpeak}
f^l_{\rm peak}= \frac{c^3}{G}\sqrt{(V_l)_{\rm max}}\approx \frac{c^3}{G}\frac{l}{2 \pi \sqrt{27} M},
\end{equation}
while, the total spectrum obtained by summing over all the multipoles, peaks at
\begin{equation}\label{eq:fpeak2}
f_{\rm peak} \approx \frac{c^3}{G}\frac{0.05}{M} \approx 10.36 \frac{M_\odot}{M}~{\rm kHz}.
\end{equation}

Then, \citet{1971ApJ...170L.105P} showed that the multipole spectra obtained in \citet{1971PhRvL..27.1466D} were associated with the $2^l$-pole normal-mode vibrations of the black-hole, excited by the gravitational-wave train produced by the in-falling body. Thus, the gravitational-wave spectrum from the peak on is governed by the emission of the vibrational energy of the black-hole driven by gravitational-wave radiation. Such vibrations are referred today to as black-hole  ``ringdown'' or ``ringing tail'' \citep{1971PhRvL..27.1466D}. 

\begin{figure}
\includegraphics[width=\hsize,clip]{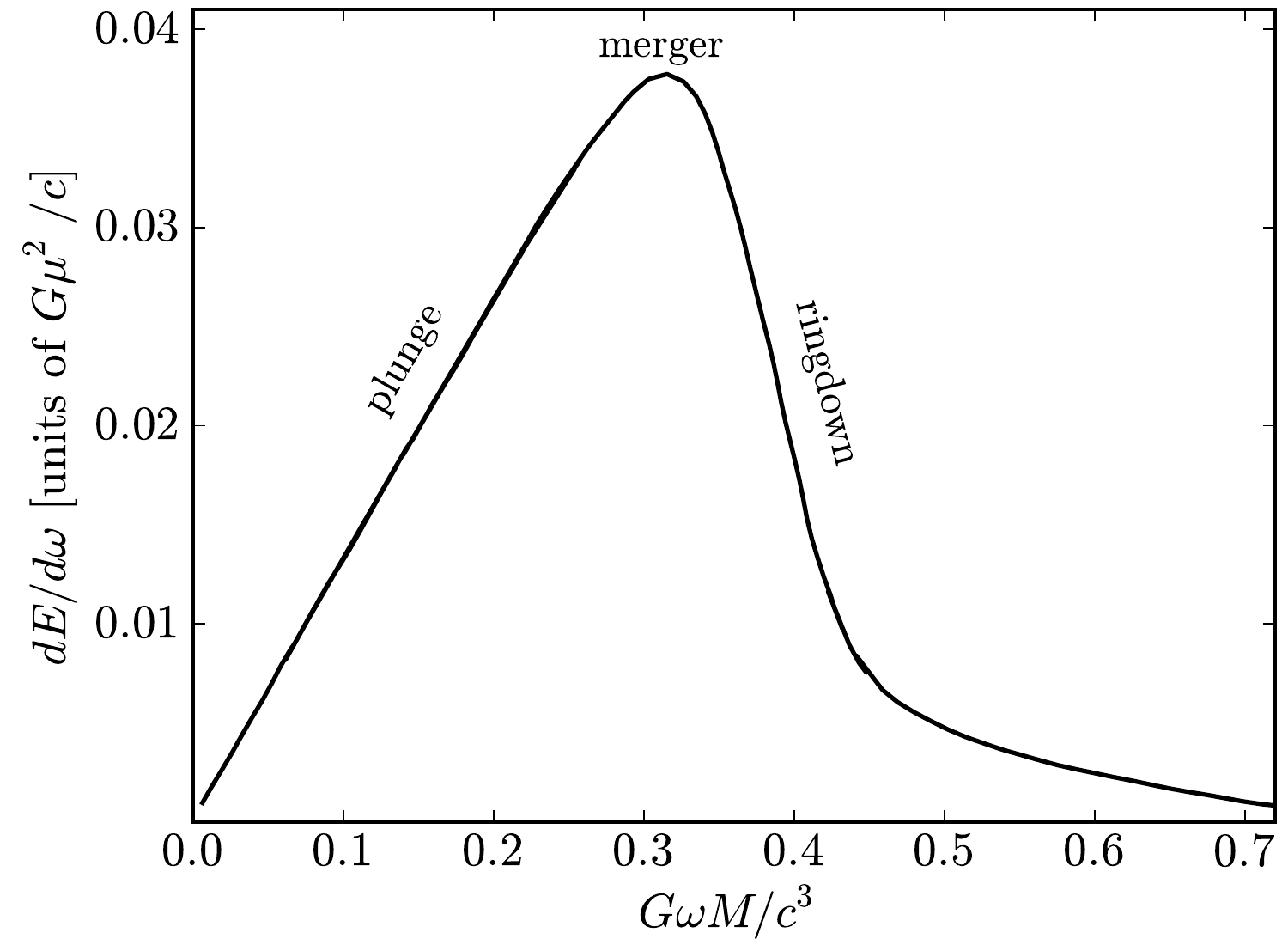}
\caption{Spectrum of the gravitational-wave radiation ($l=2$ multipole) emitted by a test-particle of mass $m$ falling radially into a black-hole of mass $M$ (in geometrical units). This figure has been adapted from the original one in \citet{1971PhRvL..27.1466D}.}\label{fig:DRPP}
\end{figure}

As it was shown in \citet{1971PhRvL..27.1466D}, the gravitational-wave spectrum in this phase has a peaked form: it first has a raising part that follows a power-law behavior which can be understood as follows. The solution of the particle radially falling onto the black-hole in linearized theory, i.e. in a flat background, and using Newtonian equations of motion which leads to \citep{1974bhgw.book.....R}
\begin{equation}\label{eq:dEdfmerger2}
\left(\frac{dE}{df}\right)_{\rm plunge} \approx 2\pi\times 0.18 \frac{G \mu^2}{c} \left(\frac{4\pi G M f}{c^3}\right)^{4/3}.
\end{equation}
The spectrum thus raises following approximately (\ref{eq:dEdfmerger2}) until it reaches a maximum at the peak frequency (\ref{eq:fpeak2}), and then it falls off following an approximate (phenomenological) form
\begin{equation}\label{eq:dEdfmerger}
\left(\frac{dE}{df}\right)_{\rm ringdown} \approx 2\pi \frac{G \mu^2}{c} \exp(-9.9 \times 2\pi G M f/c^3).
\end{equation}
The spectrum of the $l=2$ multipole radiation as obtained numerically in \citet{1971PhRvL..27.1466D} is shown in figure~\ref{fig:DRPP}. We have indicated the location of the plunge, merger and ringdown phases in the frequency-domain. It is clear that an approximate analytic formula of the spectrum can be obtained from the interpolation function
\begin{equation}
\frac{dE}{df} \approx \left[\frac{1}{(dE/df)_{\rm plunge}} +\frac{1}{(dE/df)_{\rm ringdown}}\right]^{-1}.
\end{equation}

The total energy radiated to infinity in gravitational-waves during this plunge+merger+ringdown phase is \citep{1971PhRvL..27.1466D}:
\begin{equation}\label{eq:Emerger}
\Delta E_{\rm merger} = \sum_{l\geq 2} \int df\left(\frac{dE}{df}\right)_{2^l-{\rm pole}} \approx 0.01 \frac{\mu^2}{M} c^2.
\end{equation}

From all the above we can extract three important theorems:

\begin{enumerate}

\item the final gravitational-wave frequency of the inspiral phase, $f_{\rm LSO}$, is lower than the peak frequency, $f_{\rm peak}$;

\item the energy emitted in gravitational-waves during the total inspiral phase is larger than the energy emitted in the final plunge-merger-ringdown phase;

\item the merger point can be set as the point where the gravitational-wave spectrum, $dE/df$, reaches the maximum value;

\item from the above, the frequency at merger is
\begin{equation}\label{eq:fmerger}
f_{\rm merger} \equiv f_{\rm peak},
\end{equation}
where $f_{\rm peak}$ is given by equation~(\ref{eq:fpeak}).

\end{enumerate}

\subsection{Angular momentum in the merger phase}\label{sec:2c}

As it was shown in \citet{1979ApJ...231..211D}, the energy emitted during the plunging of a test-particle into a black-hole is affected by the initial angular momentum of the particle. The total energy output in form of gravitational-wave radiation was there computed for selected initial angular momenta of the particle (which correspond to start the plunge of the particle from different orbits). It is worth to notice that in the case $J=0$, the total energy emitted approaches the numerical value obtained in \citet{1971PhRvL..27.1466D}, see equation~(\ref{eq:Emerger}), in the case of radial falling into the black-hole, as expected.

The results of the numerical integration of \citet{1979ApJ...231..211D} are well-fitted (with a maximum error of $\sim 10\%$) by the phenomenological function
\begin{equation}\label{eq:EmergerJ}
\Delta E_{\rm merger} \approx \Delta E_{\rm merger}^{J=0} [1+0.11 \exp(1.53 j)],
\end{equation}
where $j\equiv c J/(G\mu M)$ and $\Delta E_{\rm merger}^{J=0}$ is the energy radiated by a particle falling radially given by equation~(\ref{eq:Emerger}).

Thus, from the knowledge of the angular momentum at the LSO, we can infer the amount of energy emitted during the final merger phase. The energy loss during the regime II is therefore $\Delta E_{\rm merger} \approx 0.24 (\mu^2/M) c^2$, where we have used equations~(\ref{eq:Emerger}), (\ref{eq:EmergerJ}), and the fact that $j_{\rm LSO} = c J_{\rm LSO}/(G \mu M) = 2\sqrt{3}$ is the dimensionless angular momentum of a test-particle in the LSO around a Schwarzschild black-hole. 

From the amount of energy emitted in this final plunge phase, $\Delta E_{\rm merger}$ given by equation~(\ref{eq:EmergerJ}), we can estimate the angular momentum loss by the gravitational-wave emission in the final plunging, $\Delta J_{\rm merger}$, as
\begin{equation}
\Delta J_{\rm merger} \approx \frac{\Delta E_{\rm merger}}{\pi f_{\rm LSO}},
\end{equation}
which leads to $\Delta J_{\rm merger}\approx 3.81 G \mu^2/c$.

\section{Mass and spin of the formed black-hole}\label{sec:3}

In order to give an estimate of the newly formed black-hole parameters, we can use both energy and angular momentum conservation. Energy conservation implies a mass of the newly formed black-hole
\begin{align}\label{eq:mBH}
    m_{\rm BH} &\approx M - (1-2\sqrt{2}/3)\nu- \Delta E_{\rm merger}/c^2 \\
    &\approx M \beta(\nu),
\end{align}
where $\beta(\nu) \equiv \left[1-\left(1-2\sqrt{2}/3\right)\nu- 0.24 \nu^2\right]$, while angular momentum conservation leads to
\begin{equation}
J_{\rm BH} = J_{\rm LSO} - \Delta J_{\rm merger},
\end{equation}
which implies a dimensionless angular momentum of the newly formed black-hole
\begin{equation}\label{eq:alphafull}
\alpha \equiv \frac{c J_{\rm BH}}{G m_{\rm BH}^2} \approx \frac{2\sqrt{3}\nu - 3.81 \nu^2}{\beta(\nu)^{2}}.
\end{equation}
%

\section{Analysis of GW 1509014}\label{sec:4}

In order to extract more information from the signal it is necessary to make an analysis of the frequency evolution with time. From the binary evolution in the regime I, within the point-like approximation, it can be inferred the chirp mass of the binary as:
\begin{equation}
M_{\rm chirp} = \frac{c^3}{G}\left( \frac{5}{96 \pi^{8/3}} \frac{\dot{f}}{f^{11/3}} \right)^{3/5}.
\end{equation}
We fit the evolution of the frequency with time in GW 150914 which leads to $M^{\rm obs}_{\rm chirp}\approx 30.5~M_\odot$, in agreement with the LIGO analysis, $M^{\rm obs}_{\rm chirp} =30.2^{+2.5}_{-1.9}~M_\odot$, in the detector-frame \citep{2016arXiv160203840T}.

We estimate a frequency at maximum strain $f^{\rm obs}_{\rm peak}=144\pm 4$~Hz. The uncertainties are due to the resolution of the discrete Fourier transform used to obtain the spectrogram. By using the aforementioned theorem 3, we can estimate the total mass of the binary equating the observed peak-frequency to the theoretical prediction (\ref{eq:fpeak2}):
\begin{equation}\label{eq:Mth}
M_{\rm obs} = \frac{10.36~{\rm kHz}}{f^{\rm obs}_{\rm peak}} M_\odot \approx 72\pm 2~M_\odot,
\end{equation}
where we have computed it in the observer-frame, i.e. in the detector-frame. This value is to be compared with the total mass in the detector-frame obtained from the analysis based on numerical relativity templates, $M_{\rm obs}\approx 70.3^{+5.3}_{-4.8}~M_\odot$ \citep{2016arXiv160203840T}, namely, our estimate is off only within a 2.3\% of error with respect to the full numerical relativity analysis. Clearly, the total mass in the source-frame is $M=M_{\rm obs}/(1+z)$. 

From the knowledge of the chirp mass and the total binary mass, we can extract the mass-ratio of the binary. From the definition of symmetric mass-ratio we have
\begin{equation}
\nu = \frac{\mu}{M} = \left(\frac{M_{\rm chirp}}{M}\right)^{5/3}\approx 0.24\pm 0.01,
\end{equation}
which leads to a mass-ratio 
\begin{equation}
q = \frac{m_2}{m_1} = \frac{4\nu}{\bigl(1+\sqrt{1-4\nu}\bigr)^2} \approx 0.67^{+0.33}_{-0.11},
\end{equation}
which is within the numerical relativity analysis value $q=0.79^{+0.18}_{-0.19}$ \citep{2016arXiv160203840T}. Thus, we obtain individual masses
\begin{eqnarray}
m^{\rm obs}_{\rm BH,1}&=&\frac{M_{\rm obs}}{(1+q)}\approx 43.1^{+4.3}_{-7.9}~M_\odot,\\
m^{\rm obs}_{\rm BH,2}&=&\frac{q}{1+q}M_{\rm obs} \approx 28.9^{+6.3}_{-5.9}~M_\odot,
\end{eqnarray}
which agree with the numerical relativity values $m_{\rm BH,1} = 39.4^{+5.5}_{-4.9}~M_\odot$, $m_{\rm BH,2} = 30.9^{+4.8}_{-4.4}~M_\odot$ \citep{2016PhRvL.116f1102A,2016arXiv160203840T}. 

Therefore, together with the equations~(\ref{eq:mBH}) and (\ref{eq:alphafull}), we obtain a straightforward estimate of the parameters of the final black-hole:
\begin{equation}
m_{\rm BH} \approx 70^{+2}_{-2}~M_\odot,\qquad \frac{c J_{\rm BH}}{G m^2_{\rm BH}} = 0.65^{+0.02}_{-0.02},
\end{equation}
to be compared with the numerical relativity analysis $m_{\rm BH} = 62^{+4}_{-4}~M_\odot$ and $c J_{\rm BH}/(G m^2_{\rm BH}) = 0.67^{+0.05}_{-0.07}$ \citep{2016PhRvL.116f1102A,2016arXiv160203840T}.

\section{Concluding remarks}\label{sec:5}

There are two markedly different regimes in the evolution of coalescing binary black-holes: I) the inspiraling phase up to merger and II) the final ringdown phase of the newly formed black-hole. The two regimes have to be constrained by observations with comparable accuracy. In that case, our analysis shows that it is possible to extract the parameters of the system, indicated in \citet{2016PhRvL.116f1102A}, but from a much simpler analysis of the two regimes in the test-particle approximation without the need of sophisticated numerical simulations. This is quite striking since we would expect that in the real world our test-particle approximation should not be valid in nearly mass-symmetric systems like the one proposed in \citet{2016PhRvL.116f1102A} to explain GW 150914. 

The independent observational confirmation of the two regimes is indeed essential for testing the validity of this approximation, for determining the total mass of the system and the mass of each binary component, as well as the formation of the black-hole horizon. 

It is therefore unfortunate that the signal around 150~Hz occurs just at the limit of the sensitivity of LIGO, not allowing a definite characterization of regime II. Under these conditions, regime I alone is not sufficient to determine the astrophysical nature of GW 150914, nor to assess that it was produced by a binary black-hole merger leading to a newly formed black-hole.

\bibliographystyle{mnras}
\bibliography{references}

\bsp
\label{lastpage}
\end{document}